\documentstyle[sprocl,amssymb,amsmath,epsfig,calc]{article}

\def\be{\begin{equation}}
\def\ee{\end{equation}}
\def\bea{\begin{eqnarray}}
\def\eea{\end{eqnarray}}

\newcommand{\kms}  {  {\rm km  \ s}^{-1}	}

\newcommand{\hMpc} {  h^{-1}{\rm Mpc}  		}
\newcommand{\M}    {  {{\cal M}	       }	}

\newcommand{\ltsima} {$\; \buildrel < \over \sim \;$}
\newcommand{\gtsima} {$\; \buildrel > \over \sim \;$}
\newcommand{\simlt}  {\lower.5ex\hbox{\ltsima}}
\newcommand{\simgt}  {\lower.5ex\hbox{\gtsima}}

\begin{document}

\title{A NEW LARGE CATALOG OF GROUPS OF GALAXIES \\
IN THE SOUTHERN GALACTIC HEMISPHERE}

\author{ ROBERTO TRASARTI--BATTISTONI}

\address{Theoretische Physik, Ludwig-Maximilians-Universit\"at,\\ 
Theresienstr. 37, D-80333 M\"unchen, Germany\\
email:\;\;roberto@stat.physik.uni-muenchen.de}


\maketitle\abstracts{
We present a large catalog of $N_G \approx 200$ loose groups of galaxies
selected from the Perseus--Pisces redshift Survey (PPS)
in the Southern Galactic Hemisphere.}
%

Galaxy groups can be regarded as systems intermediate between galaxies and
galaxy clusters. 
They provide constraints on cosmological models through two different routes:
as galaxy systems, through their internal properties
(velocity dispersion $\sigma_v$, radius $R$, etc.);
as LSS tracers, through their ``external'' properties
(abundance, clustering, fraction of grouped galaxies, etc.).
Loose groups of galaxies are moderate density enhancements ($\delta n/n \sim 10$-$100$)
consisting of few members ($N_{\rm mem} \sim 3$-$30$, typically 5).
By their very nature, they are then
(1) very numerous, thus suitable for statistical studies, but
(2) somehow elusive to define, thus requiring care in the identification procedure.

Several large catalogs of loose groups 
are nowaday available from the $m_{lim}=15.5$ redshift surveys CfA2Slice and CfA2North 
in the Northern Galactic Hemisphere (Ramella et al. 1989; Ramella et al. 1997)
but none of them covers the region of the Perseus--Pisces Survey (PPS)
($-1^h.50 \le \alpha \le +3^h.00$, $ 0^o\le \delta \le 40^o$, $m_{lim}=15.5$
for the sample PPS2 considered here.)
Therefore, our target is to build a new group sample which is at the same time:
(1) large, internally homogeneous, and independent from other samples
(2) objectively defined, and as homogeneous as possible to the other samples.
Groups are then identified with the redshift--space adaptive 
Friends--Of--Friends (FOF) algorithm of Huchra \& Geller (1982).
The two links $D_L$ and $V_L$ are normalized by
$D_0=0.231 \ h^{-1}{\rm Mpc}$ and $V_0=350 \ {\rm km  \ s}^{-1}$
at $cz_0 = 1000 \ {\rm km  \ s}^{-1}$,
and then scaled with $cz$ 
to compensate for the effect of $m_{\rm lim}$.

\begin{figure}
\begin{center}
\epsfig{figure=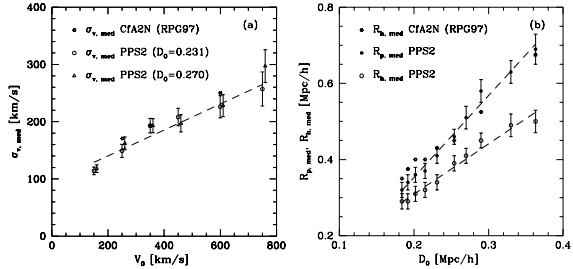,height=3.7cm,bbllx=50,bblly=430,bburx=560,bbury=700}
\caption{Group properties (medians) vs FOF links normalization.
(a) Median velocity dispersion $\sigma_v$ vs velocity link $V_0$;
(b) Median harmonic radius $R_h$ and 
pairwise inter--group member separation $R_p$ vs spatial link $D_0$.
(Values for CfA2 North from Ramella et al. 1997.)}
\label{fig:plots}
\end{center}
\end{figure}


The galaxy luminosity function (LF) appropriate for PPS 
($\alpha=-1.15 \pm 0.15$, $M_*=-19.3 \pm 0.1$, Trasarti--Battistoni 1996;
$\phi_*=0.02 \ h^{3}{\rm Mpc}^{-3}$, Marzke et al. 1994),
and the adopted $D_0$, together yield a density threshold $\delta n/n \approx 180$.
This is rather different from the desired $\delta n/n \approx 80$ 
appropriate for the other samples, mostly due to their different LF normalization $\phi_*$
since $(1+\delta n/n) \propto \phi_*^{-1} D_0^{-3}$.
We then consider several other link normalizations.
(The effects on group clustering
are discussed in Trasarti--Battistoni et al. 1997.)

Fig.\ref{fig:plots} shows that 
the internal (median) properties of FOF groups
are very directly related to the normalization of the FOF links adopted for the group catalog.
Further analysis (Trasarti--Battistoni 1996, 1997) suggests that, 
for a given normalization, the internal properties of FOF groups at different $cz$
are also simply related to the value of the FOF links at the same $cz$.
However (Table 1), for given FOF normalizations and scalings, 
the global properties of groups in PPS 
and in the directly comparable samples show good agreement.

{
\begin{table*}
\label{tab_PPSCfA}
\begin{center}
\caption[]{Group properties in PPS2 and in CfA2 (medians, and global values)}
\begin{tabular}{l l l l l l l}
\hline
Galaxy sample   &$\sigma_v$ &$R_h$   &$t_{cr}$   &$\M_{vir}$ &$N_G/\omega$ &$f_{gr}$\\
         &$\kms$     &$\hMpc$ &$H_0^{-1}$ &$h^{-1}\M_\odot$ &${\rm sr}^{-1}$  & \\ 
\hline
$D_0=0.231 \ \hMpc$ 

              &      &        &           &       &                    &    \\
PPS2	      &194   &0.34    &0.22       &1.44$\times 10^{13}$ &2.5$\times 10^{2}$  &0.35\\
CfA2 North    &192   &0.40    &0.21       &1.23$\times 10^{13}$ &3.4$\times 10^{2}$  &0.40\\
         &           &        &           &       &                    &    \\
$D_0=0.270 \ \hMpc$ 
              &      &        &           &       &                    &    \\
PPS2	      &193   &0.41    &0.26       &1.90$\times 10^{13}$ &2.6$\times 10^{2}$  &0.41\\
CfA2 Slice(s) &215   &0.41    &0.22       &2.57$\times 10^{13}$ &3.0$\times 10^{2}$  &0.44\\
\hline
\end{tabular}
\end{center}
\end{table*}
}


\begin{thebibliography}{99}

\bibitem{HG82}
Huchra  J.P., \& Geller M.J., 			1982, ApJ   257, 423

\bibitem{MHG94}
Marzke R.O., Huchra J.P.,\& Geller M.J.,  	1994, ApJ   428,  43

\bibitem{RGH89}
Ramella M., Geller M.J., \& Huchra J.P., 	1989, ApJ   344,   57

\bibitem{RPG97}
Ramella M., Pisani A., \& Geller M.J.,		1997a,  AJ   113, 2

\bibitem{TB96}
Trasarti--Battistoni R., 			1996,
Ph.D Thesis, Universit\`a di Milano
\bibitem{TBIB97}
Trasarti--Battistoni R., et al. 		1997, ApJ 475, 1 
\bibitem{TB97b}
Trasarti--Battistoni R.,			1997, A\&A Suppl. submitted



\end{thebibliography}
\end{document}